\documentclass[usenatbib,fleqn]{mn2e}
\usepackage{amsmath}
\usepackage{epsfig}

\newcommand{\apj}{ApJ}
\newcommand{\mnras}{MNRAS}
\newcommand{\aap}{A\&A}

\newcommand{\araa}{ARA\&A}
\newcommand{\nature}{Nat}

\title[SCUBA observations of dark GRB hosts]{SCUBA observations of the host galaxies of four dark gamma-ray bursts}

\author[V.E.~Barnard et al.]
{V.E.~Barnard$^1$, 
A.W.~Blain$^{2}$, 
N.R.~Tanvir$^3$, 
P.~Natarajan$^4$, 
I.A.~Smith$^5$,\newauthor
R.A.M.J.~Wijers$^6$, 
C.~Kouveliotou$^7$, 
E.~Rol$^8$, 
R.P.J.~Tilanus$^9$, 
and P.~Vreeswijk$^8$ 
\\
\\
$^1$Cavendish Astrophysics, University of Cambridge, Cambridge, CB3 0HE, UK \\
$^2$Dept. of Astronomy, Caltech 105-24, Pasadena, CA 91125, USA\\
$^3$Dept. of Physical Sciences, University of Hertfordshire, Hatfield, AL10 9AB, UK\\
$^4$Dept. of Astronomy, Yale University, New Haven, CT 06250-8181, USA\\
$^5$Dept. of Physics and Astronomy, Rice University MS 108, Houston, TX 77005-1892, USA\\
$^6$Dept. of Physics and Astronomy, SUNY, Stony Brook, NY 11794-3800, USA\\
$^7$NASA Marshall Space Flight Center, SD-50, NSSTC, 320 Sparkman Drive, Huntsville, AL 35805, USA\\
$^8$Astronomical Insitute `Anton Pannekoek', University of Amsterdam and Center for High-Energy Astrophysics, \\
~Kruislaan 403, 1098 SJ Amsterdam, The Netherlands\\
$^9$Joint Astronomy Centre, 660 N, A'ohoku Place, Hilo, HI 96720, USA\\
}

\begin{document}
\maketitle

\begin{abstract}
We present the results of a search for submillimetre-luminous host galaxies of
optically dark gamma-ray bursts (GRBs) using the Submillimetre Common-User 
Bolometer Array
(SCUBA) on the James Clerk Maxwell Telescope (JCMT).  We made
photometry measurements of the 850-$\mu \rm{m}$ flux at the location
of four `dark bursts', which are those with no detected optical afterglow despite rapid deep searches,
and which may therefore be within galaxies containing substantial amounts of dust.  
We were unable to detect any individual source significantly. Our results are consistent with predictions for the host galaxy population as a whole, rather than for a subset of dusty hosts.  This indicates that optically dark GRBs are not especially associated with very submillimetre-luminous galaxies and so cannot be used as reliable indicators of dust-enshrouded massive star-formation activity.  Further observations are required to establish the relationship between the wider GRB host galaxy population and SCUBA galaxies.
\end{abstract}


\section{Introduction}
\label{sec:intro}

\subsection{Gamma-ray Bursts}
\label{subsec:grbtheory}

It is generally now accepted that the afterglow emission resulting
from (long-duration, soft-spectrum) gamma-ray bursts (GRBs) can be
explained by an ultra-relativistic shock-wave expanding into a
surrounding medium \citep{mesrees,GRBreview,mesrev}.  The
precise nature of the progenitor systems is not a settled issue. 
The two 
most popular theories both involve stellar remnants: the collapsar/hypernova model, in which a single
massive progenitor star undergoes core collapse
\citep{woosley,microquasar}; and the binary merger theory,
in which two massive stellar remnants, such as neutron stars, merge \citep{lattimer,binary}. Both of these scenarios may be able to explain the energetics of the explosion that produces
the GRBs, particularly if they are beamed, since the energies of GRBs are comparable to those involved in the formation of typical stellar-progenitor black holes.

Recently, the collapsar/hypernova model has gained support from
three sets of observations: firstly, it was confirmed that the
positions of some GRBs accurately localized by the observation of optical
and/or radio afterglows, were found to be within star-forming
regions of their host galaxies \citep{offset}, which themselves are frequently starburst galaxies \citep{sokolov}.    This is supported by X-ray determinations of the HI column density along the line of sight to GRBs, which is consistent with their residing in GRBs \citep{galama}.  During their creation,
it is likely that the massive stellar remnants required in the binary merger scenario would
receive a substantial `kick' velocity, so that the merger event causing the GRB would take place outside of the star-forming region of the host galaxy \citep{galama}. Also, the delay required between formation of the remnants and their merger may well be long enough for star formation to have ceased in the host, suggesting that the host galaxy would no longer be luminous.  Hence the position measurements of \citet{offset} support the single massive progenitor theory.
Note though, that one should be 
wary of a potential selection effect: the detection of a GRB afterglow requires a certain minimum density in the surrounding interstellar medium (ISM), and so 
GRBs may only be identifiable when they occur inside 
galaxies \citep{chevalier}.  

Secondly, in a few
cases, optical afterglows of GRBs have been seen to contain a
secondary brightening in flux after a few weeks \citep{sn980425,sn980326,reichart,
galamaII,000911SN}. This flux increase, though often not very 
significant, has been attributed to a supernova 
occurring simultaneously with the GRB, although alternative 
ideas have been postulated, such as dust echoes \citep{esin} 
or interactions of the GRB shock wave with wind-driven density structures in the surrounding ISM 
\citep{EnricoIII}. 
Supernovae associated with GRBs are compatible with the single progenitor collapsar/hypernova 
model 
rather than the merger hypothesis. 

Finally, the detection of iron features 
in the X-ray afterglows suggest the presence of an iron-enriched ISM surrounding the GRB progenitor \citep{amati,piro,yoshida}.  
This high iron mass is consistent with ejecta from a massive stellar progenitor.

If the hypernova theory is the correct explanation for long-duration GRBs, then there should be a direct link between GRBs and high-mass star formation activity. Since the gamma-ray emission from the initial explosion is not 
attenuated by dust, and can be detected from high redshifts, 
GRBs should be unbiased tracers pointing to star 
formation activity wherever 
massive stars are living and dying
\citep{krumholz,BlainPriya,b32,010222,EnricoII}.

\subsection{SCUBA galaxies}
\label{subsec:SCUBAgals}

Since the commissioning of the SCUBA instrument \citep{wayne} in 1997 
on the JCMT on Mauna Kea, a new era of submillimetre (submm) cosmology has been possible (see, e.g. \citealt{bsik,s38}).  In particular, the
discovery of a substantial population of dust-enshrouded `SCUBA
galaxies' has launched a debate regarding observational estimates of
the global star formation rate in the Universe.  SCUBA traces the
interstellar dust in galaxies (with a temperature of the order of tens of degrees Kelvin),
which may be heated by the UV light emitted by OB stars
and/or a hosted AGN. The relative contributions  of starlight
and AGN to dust-heating in a particular galaxy are hard to determine precisely, but studies using \emph{Chandra} to observe
hard X-rays from submm-selected galaxies suggest that about
20 per cent of the sample contain a detectable AGN
\citep{bautz,fabian,chandra8mjy}. 

Therefore, it is likely that the luminous dust emission from
SCUBA galaxies is powered predominantly by star formation. In this
case, because of the large 
bolometric luminosities of the SCUBA galaxies (at least $10^{12} L_{\sun}$), they make a
substantial contribution to overall star formation activity in the high-redshift 
universe,
comparable to or greater than that of optically-selected galaxies
\citep{bsik}.  Thus, they are good candidates to be hosts of many GRBs, if GRBs are indeed associated with the death of high-mass stars.  The optical properties of SCUBA galaxies, where they are well-located, indicate that most are very faint, with $R\,>\,25$ \citep{s38}, not dissimilar to the GRB host galaxy population as a whole \citep{970828}. 
It is likely that about 10--20 per cent of the star-formation activity in the high-redshift universe takes place in submm galaxies brighter than 2\,mJy.  As a result, perhaps 1 in 5 GRBs should reside in such objects.  This point is examined further in Section \ref{sec:results}.

\subsection{Submm observations of GRBs}

This paper describes our submm SCUBA observations of the
host galaxies of four GRBs.  To increase the chances of finding dusty
SCUBA galaxies in this pilot program, we selected GRBs which were `dark'
in the sense that their afterglows were undetected in the optical 
despite deep, rapid searches.  Estimates vary, but optical
afterglows have been searched for and not found in roughly 30-50 per cent of
GRBs with X-ray afterglows. In many cases the non-detection may
simply be because the searches were not deep or rapid enough (Galama
\& Wijers 2001), but in some instances it is clear that the optical afterglows
are genuinely underluminous (Groot et al. 1998, 
van Paradijs et al. 2000, Lazzati et al. 2001a, Ramirez-Ruiz et al. 2002).
An obvious possibility is that these bursts are heavily obscured by dust.
Since GRBs are expected to destroy any dust in their vicinity
\citep{waxdrain,fruchter,reichyost,bram}, the obscuration would have to be due to dust elsewhere
along the line of sight, as might well be expected in very dusty submm-bright galaxies.

Note that we did not attempt to observe GRB afterglows in the submm. 
Target of Opportunity programs to observe GRB afterglows are underway at 
the JCMT and IRAM 30-m telescopes
\citep{SmithI,SmithII,010222}. 
Our observations took place long enough after the initial explosion
for only the host galaxy emission to be detectable.

There have been two previous SCUBA detections of a GRB host galaxy.  The first was found serendipitously during a submm 
afterglow search \citep{010222} for a GRB that had an optical afterglow (GRB 010222). This suggests that the GRB was located in a luminous dusty galaxy, but either in a relatively
dust-free region or near the edge of the galaxy, so that the explosion could clear
out any dust along the line of sight.  Also, 
\citet{b32} reported the results of a successful targeted search for 
submm emission from  GRB 000418.  This too had an optical transient identification.  Another relevant detection was that of a GRB host in the radio (GRB 980703; \citealt{BergerIII}), whose flux was claimed to be caused by a large star-formation rate rather than AGN activity, based on optical spectroscopy and the absence of radio variability. However, a deeply embedded AGN within a dust- and gas-rich galaxy could plausibly contribute the emission from this galaxy, especially as the radio source is located very near to the nucleus of the host galaxy.

\section{Observations}
\label{sec:obs}

Observations were made using the 850-$\mu \rm{m}$ photometry pixel on
the SCUBA array \citep{wayne}.  They are summarized in Table
\ref{obstable}.  Simultaneous observations were made with the 450-$\mu
\rm{m}$ photometry pixel, but weather conditions and/or array
noise were never good enough to yield useful data.  Observations were made in the standard photometry mode, using a
7 Hz, 60\arcsec\ `chop' in azimuth to provide a blank-sky reference
and a further telescope `nod' to produce a measure of any sky gradient.  Each
set of photometry observations takes 18 seconds.  Column 3 of Table 1 shows the integration time on-source for each GRB.  Pointing was checked
regularly and was always better than 2.5\arcsec, much smaller than the primary beam of the array of 14\arcsec at 850 $\mu \rm{m}$.  Seeing was monitored, especially around dawn and dusk, and
data was only taken when the seeing was less than 1\arcsec.  Sky
transparency was calculated by interpolating between regular skydip
values, though the situation was monitored more frequently with data
from the JCMT's water vapour monitor.  The average sky opacity at 850 $\mu \rm{m}$ during each source's observations, $\tau_{\ 850 \mu \rm{m}}$, is given in column 4 of table 1.  For flux
calibration, identical observations were made of the planets Mars and
Uranus, and where necessary the secondary flux calibrator CRL 618.  We
found no significant deviations between the observed and expected
fluxes of these standard objects.

Unfortunately, since the observations were made, two of the GRBs in our sample have been found to have been wrongly located: GRBs 001109 and 001025.  In the case of GRB 001109, the radio-located position 
which we observed is now thought to correspond to a faint 
constant radio source \citep{BergerII}, and thus to be a mis-identification 
of the GRB afterglow. We detected a net positive flux of $(1.89 \pm 1.40)$ mJy from this source, consistent with its likely identification as a high-redshift star-forming galaxy \citep{BergerII}.  For GRB 001025, \emph{XMM} error box S1 was observed, since it contained a candidate host galaxy (Hjorth priv. comm.), but subsequently Hurley (priv. comm.) has calculated that this error box lies outside a revised \emph{IPN} annulus.  Hence the results of the observations of both these objects are not included in the rest of this paper.

\begin{table}

\caption{Photometry observations log.  See notes about GRB 001025 and GRB 001109 in section \ref{sec:obs}.} 

\begin{tabular}{lccccc}\hline
Source name & Date of obs. & Integ. time (s) & $\tau_{\ 850 \mu \rm{m}}$\\\hline
GRB 970828 & 24/04/01 & 900 & 0.23 \\
GRB 981226 & 03/10/01 & 2052 & 0.27 \\
& 24/11/01 & 2115 & 0.27 \\
GRB 990506 & 08/05/01 & 2250 & 0.26 \\
GRB 000210 & 01/09/01 & 1800 & 0.24 \\
& 03/10/01 & 2250 & 0.27 \\
GRB 001025 & 05/10/01 & 2250 & 0.30 \\
GRB 001109 & 08/05/01 & 1350 & 0.26 \\
& 03/10/01 & 2700 & 0.26\\\hline
\end {tabular}
\label{obstable}
\end{table}  

\section{Data Reduction}
\label{sec:data}

The data were reduced using the standard Starlink SURF procedures
\citep{surf}.  Particular care was taken over the removal of
atmospheric noise, since the expected low fluxes of the sources make
this an important factor.  This sky-noise removal is possible using
the other bolometers on the array, assuming that they are pointing at
the blank sky.  Following \citet{kate'squasars}, the median value of the
signal from the reference bolometers was used rather than the mean.  
Usually the
reference bolometers used are the inner ring on the 850-$\mu \rm{m}$
array.  For the October 2001 observations (see table \ref{obstable}),
however, the whole array suffered from elevated noise.  
Particular care was taken to choose only those bolometers with
normal noise levels for this final sky-removal stage, and the
bolometer noise values were measured far more frequently than usual.
The results presented have been clipped at the 3$\sigma$ level, though
this generally had little effect on the final results.

\section{Results and Discussion}
\label{sec:results}

Table \ref{restable} shows the overall weighted mean at 850 $\mu \rm{m}$ for each of the
four reliable sources observed.  The overall, weighted mean flux for all four
sources is -0.37 $\pm$ 0.82 mJy.

\begin{table}
\centering
\caption{Coadded 850-$\mu \rm{m}$ results for the four objects observed.  Redshifts $z$, and R magnitudes, where determined, for candidate host galaxies are shown here (\citealt{970828} (GRB 970828), \citealt{hollandII} (GRB 981226), \citealt{bloomAJ} (GRB 990506) and \citealt{p15} (GRB 000210)).}

\begin{tabular}{cccc}\hline
Source  & Flux (mJy) & $z$ & R mag\\\hline
GRB 970828 & 1.26 $\pm$ 2.36 & 0.96 & 25.5\\
GRB 981226 & -2.79 $\pm$ 1.17 & - & 24.3\\
GRB 990506 & -0.25 $\pm$ 1.36 & 1.3 & 24.8\\
GRB 000210 & 3.31 $\pm$ 1.54 & 0.85 & 23.5\\\hline

\end {tabular}
\label{restable}
\end{table}

\begin{figure}
\centering
\vspace*{9cm}
\leavevmode
\includegraphics{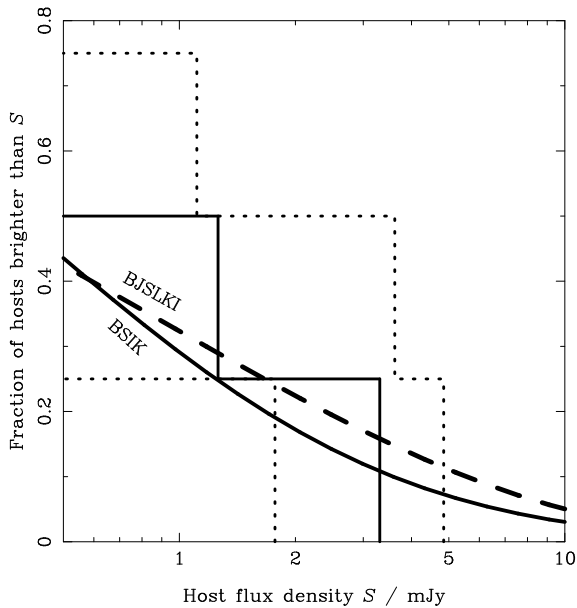}
\caption{Comparison between results and predictions\ \citep{EnricoII}.  The smooth curves show the fraction of all GRB host galaxies expected to exceed a 
given 850-$\mu$m flux in two current models -- for a pure luminosity evolution (BSIK: \citealt{bsik}) and an hierarchical model respectively (BJSLKI: \citealt{bjslki}).  Stepped lines show results from this pilot study - see text.
}

\label{figpred}
\end{figure}

Models for the evolution of the star formation rate in dusty galaxies can 
be used to predict the likely 850-$\mu
\rm{m}$ fluxes of GRB hosts \citep{bsik,bjslki,BlainPriya,EnricoII}, as 
shown in Fig.\,\ref{figpred}. We assume that:  
dust in SCUBA galaxies is predominantly heated by high-mass stars; 
GRB rates are tied to the rate of formation of high-mass stars; and
most high-redshift star formation activity is
enshrouded by dust. Results for two 
models based on infrared and submm data are presented, a simple parametric model of the 
evolution of low-redshift galaxies (BSIK: \citealt{bsik}) and a model based on luminous hierarchical 
merging of galaxies (BJSLKI: \citealt{bjslki}). These models both provide a good 
representation of the evolution of dusty galaxies, measured in other observations. 
About 20 per cent of \emph{all}
GRB host galaxies are expected to have fluxes above the SCUBA confusion limit 
for detection of
2\,mJy; about 10 per cent are expected to have 
fluxes greater than 5\,mJy. In general, we predict an average
across \emph{all} sources of 0.9\,mJy. The solid histogram 
shows the result of our observations, and the dotted histograms  
define the $\pm 1 \sigma$ errors in flux.

It is clear from Fig.\ \ref{figpred} that the errors on our results are
consistent with the predictions of the galaxy evolution
models, and that the overall mean flux is within $2 \sigma$ of the
expected value.
GRB 000210 is clearly our best candidate for a detection, 
but only at a significance of about  $2\sigma$. 
However, observations of GRB 000210
suffered from the raised array noise discussed in section
\ref{sec:data}, and from its low elevation from Mauna Kea. Hence, the 
atmospheric noise for GRB 000210 was greater than for our other sources.

Note that we purposefully chose a selection bias that should have
increased our chances of finding dusty GRB hosts, by observing  `dark bursts'.  Since we did not see any increase in our detection
rate, alternative explanations are necessary for the lack of optical
afterglow detections. We now look briefly at the four reliable GRBs in our
sample in more depth.

\subsection{Individual Sources}

\subsubsection{GRB 981226}

Three early candidates for the optical afterglow of GRB 981226 
\citep{gcn173,gcn172,gcn177} were rejected upon the
location of a radio afterglow \citep{981226VLA}. The limit on the optical afterglow is therefore R $>$ 23.5 at 10 hours after the alert \citep{lindgren}.  Later HST/STIS and
VLA imaging has located the probable host galaxy \citep{hollandII}.

The multi-wavelength afterglow emission from GRB 981226 provides some clues. The X-ray
afterglow was found to have a double-peaked structure followed by a rapid decay \citep{981226BS}.  Also, \citet{981226VLA} noted a rapid decline in the radio afterglow. Taken together,
the X-ray, optical and radio afterglow behaviour may all be explained
by a complicated density structure in the ISM around the
GRB.  In particular, a cavity in the ISM density would
explain the rapid X-ray and radio decays, and would predict a similarly
fast optical decay, meaning that the optical afterglow searches were too
slow/shallow to detect the emission. Such a structure could be produced
by the mass loss phases that massive stars (which are of course
plausible GRB progenitors) are known to go through in their post main
sequence lives \citep{chevalier,EnricoIII}.  Hence, there may be no
need to infer the presence of dust around this GRB.  Other suggestions
for rapid decays have been proposed which also do not rely on the presence of dust, such as the effects of jet structure \citep{jets,panaitescu}.

Our non-detection is consistent with the brief
weather-affected SCUBA afterglow search for this GRB by \citet{SmithI}.

\subsubsection{GRB 990506}

GRB 990506 appears to have some similarities with GRB 981226.  Again it
had a very rapidly decaying radio transient \citep{t11}.  The optical transient was not detected at R = 19 after 1 hour \citep{zhu}, nor to R = 23.5 after 11 hours \citep{pedersen}.  The rapid decay of the radio afterglow again suggests that a non-dusty effect, local to
the burst, may explain the absence of an optical afterglow.  Optical searches for the
host have identified it as a very faint and compact galaxy
\citep{offset}.

\subsubsection{GRB 970828}

GRB 970828 was dark to a depth of R = 23.8 \citep{groot,970828} in
observations taken from 4\,hrs after the burst.  A radio flash was observed by the VLA
\citep{970828} and has been interpreted as reverse shock emission. No
conventional radio afterglow was detected.  Subsequent Keck and \emph{HST}
observations of the location of the radio flare revealed an
interacting three-component host \citep{970828}, with a possible
identification of the GRB location in a dust lane between the two
brightest components.  In calculations considering both jet and
spherical models for the GRB shock geometry, Djorgovski et al. (2001)
conclude, based on the X-ray afterglow flux,
that a single typical Giant Molecular Cloud could provide all
the extinction necessary to fit the upper limits to the optical
afterglow flux density.  They also note that the Keck and \emph{HST} images indicate that the two host galaxy components on either side of the GRB location are both slightly but not
highly reddened, suggesting a low total dust mass in the system.\footnote{There are however indications from observations of galaxies known to be very luminous at submm wavelengths \citep{ivison} that regions where dust emission is strong may not correlate with regions which at optical wavelengths appear to be reddened by strong extinction.  Hence the lack of obvious extinction in optical images does not rule out the presence of a large amount of illuminated and heated dust.}  Our positive but not significant measurement supports the hypothesis that in this case at least, it was
not galaxy-wide dust that caused the obscuration of the optical
afterglow, but rather a localized cloud or clouds of dust along the line-of-sight, consistent with the X-ray results.  Of course, SCUBA's resolution is such that our observations are sensitive only to the total (illuminated) dust in the entire system.

\subsubsection{GRB 000210}

GRB 000210 represents our most likely host galaxy detection.  Both X-ray and radio transients were found for this source, but no optical transient (R $>$ 23.5 at 12.4 hours after the burst, \citep{g15,p15}). The X-ray transient did not display the rapid decays found for GRBs 981226 and 990506, leading Piro et al. (2002) to reject the no-dust hypotheses discussed above.  Instead they conclude that the most likely scenarios are either obscuration of the optical transient by a clumpy local environment, or line-of-sight obscuration by the whole host galaxy, either of which is allowed by our findings.  

\section{Conclusions}
\label{sec:conc}

From our small sample of four reliably-identified dark GRBs, we find that
the `dark bursts' do not preferentially select dusty host galaxies with
very significant amounts of star formation. Looking at some members of our
sample, we can explain the lack of optical transients for other reasons.
It seems likely that to characterise the optically dark bursts as a
physically distinct population of GRBs would be misleading. In
each case different circumstances due to a combination of the observing
conditions and the physical conditions at the location of the GRB could
give rise to the lack of an optical afterglow.  If, instead, we view
the GRBs included here as four examples of the overall population, we find
that the results agree with the distribution predicted
assuming that GRBs tracew high-mass star formation.  Our
lack of a single strong detection implies that no more than 20 per cent of
GRB hosts are submm galaxies detectable at a flux density brighter than
5\,mJy using SCUBA. In interpreting the results, however, the small
sample size inevitably makes it hard to draw any solid conclusions.

Combining our three most significant results with the observations of
\citet{SmithI,SmithII}, we find that 11 GRB locations have been observed
with SCUBA to noise levels $<$ 2mJy, and in all but two (GRB 980329,
\citealt{SmithI} and GRB 000210, this paper) detection of submm emission from a host galaxy was ruled
out. The two GRB hosts detected by SCUBA \citep{010222,b32} had afterglows with optical transients.  Therefore, while some GRBs definitely are located in
dusty galaxies, the route to selecting these GRBs on the basis of their
afterglow data is not yet clear.  Since GRBs can only clear dust out to less than 100 pc distance \citep{fruchter}, it would be surprising if optically-selected samples of GRBs are not generally biased against dusty hosts.  

However, as noted above, there may be alternative explanations for the optical faintness of some afterglows beyond location in a dusty host galaxy.  Our results indicate that radio-located optically-dark bursts seem not to be reliable indicators of luminous,
dusty host galaxies.  It may be that the physical
conditions of the ISM in the densest star forming regions
are incompatible with the generation of intense radio emission
from GRB shocks.  In that regard, it would ultimately be
interesting, and certainly possible in the SWIFT era, to
study a sample of hosts of purely X-ray selected GRBs,
with the hope that the prompter X-ray emission would be 
less affected by the wider environment of the progenitor.

From the SCUBA observations alone, we cannot separate the possibility that
there is little dust in these systems from the possibility that there is a
lot of dust but insufficient UV photons to make it glow brightly.
Alternatively, the dust may be heated to high temperatures and so
cannot be detected by SCUBA at all.
The predicted flux of a $5 \times 10^{12} \rm{L}_{\sun}$ galaxy at $z = 1$
decreases from 20\,mJy to 0.08\,mJy as the dust temperature increases
from 20\,K to 80\,K (see Fig.\,5 in \citet{blain_rev}). Typical dust
temperatures for submm-selected objects
are thought to be 40\,K \citep{ivison2000}, but a temperature $\sim 60$\,K
would be hot
enough to prevent such a galaxy being detected above the 2\,mJy confusion
limit of SCUBA.  However, optical colours of the host galaxies of GRBs
970828, 981226 and 000210 \citep{970828,981226VLA,p15} show only modest
reddening, to the optical depths that can be probed, and thus perhaps
imply little dust overall in the hosts. Ideally,
high-resolution observations to locate dust-enshrouded star-formation
activity with respect to the location of GRBs need to be made.

The results presented here suggest that a more extensive deep submm survey
of GRB hosts is necessary to investigate their far-infrared and submm
properties in detail, and this is now underway at the JCMT (Barnard et
al., in prep.).

\section*{Acknowledgements}

VEB is supported by a PPARC studentship.  We
thank Daniel Mortlock, Kate Isaak, Enrico Ramirez-Ruiz and Dave Green
for helpful discussions and we thank Ian Smail for a useful referee's report.  We thank J. Greiner for keeping his useful website up to date (http://www.aip.de/$\sim$jcg/grbgen.html) as well as all the observers and staff at the JCMT who carried out flexible-scheduled observations on our behalf.

\end{document}